\documentclass[aps,prl,twocolumn,dvipdfmx]{revtex4-2} %floatfix,
\usepackage[dvipdfmx]{graphicx} 
\usepackage[dvipdfmx]{color}
\usepackage{amsmath}

\begin{document}
\title{Valence Bond Crystal Ground State of the 1/9 Magnetization Plateau in the Spin-1/2 Kagome Lattice}

\author{Katsuhiro Morita}
\email[e-mail:]{katsuhiro.morita@rs.tus.ac.jp}
\affiliation{Department of Physics and Astronomy, Faculty of Science and Technology, Tokyo University of Science, Chiba 278-8510, Japan}

\begin{abstract}
We investigate the ground state of a spin-1/2 kagome antiferromagnet at the 1/9 magnetization plateau, focusing primarily on six types of valence bond crystal (VBC) distortions. 
Among six types of VBC distortions, type~1 consistently exhibits the lowest ground-state energy.
Analysis of the second derivative of the energy with respect to the distortion strength $J_{\rm d}$ using exact diagonalization reveals that, as the system size approaches the thermodynamic limit, the type~1 VBC state remains stable down to $J_{\rm d} = 1$, corresponding to the undistorted kagome lattice.
Perturbation theory further supports the stability of the type~1 VBC state, providing energy values that agree with exact diagonalization results within 3\%.
We also explore the possibility that other types of VBC distortions could become the ground state, but type~1 VBC still exhibited the lowest energy.
These findings suggest that the ground state of the 1/9 plateau in the kagome lattice is the type~1 VBC state.
\end{abstract}

\maketitle

The kagome lattice has attracted significant attention in the study of frustrated quantum spin systems due to its unique geometric properties~\cite{KLR}. In particular, the spin-1/2 kagome antiferromagnet has been a central focus of both theoretical and experimental research because of its potential to host exotic quantum states such as spin liquids~\cite{KLZ2-1,KLZ2-2,KLZ2-3,KLU1-1,KLU1-2,KLU1-3,KLU1-4,KLU1-5} and valence bond crystals (VBCs)~\cite{KLVBC1,KLVBC2,KLVBC3}, as well as the presence of  magnetization plateaus~\cite{KLMH1,KLMH3,KLMH4,KLMH5,KLMH7,KLMH8,KLMH9,YCOHC7} and  plateau-like anomalies~\cite{KLMH2,KLMH6,Ameltkagome1,Ameltkagome2,Ameltkagome3}. 
In particular, the 1/9 magnetization plateau has recently attracted significant attention, partly due to its recent observation in experiments~\cite{YCOHB1,YCOHB2,YCOHB3,YCOHB4,YCOHB5}. 

Theoretically, two main proposals have been made regarding the nature of  the 1/9 plateau. 
One is a VBC state, where localized spin singlets form with broken symmetry~\cite{KLMH5,KLMH8}. 
The other is a $Z_3$ spin liquid state with topological order~\cite{KLMH4,1-9Z3}.
Despite these theoretical advances, it remains unclear which of these states is realized in the system.
Furthermore, it is also uncertain whether there are other possible candidates.

On the experimental side, recent experimental studies have reported the observation of the 1/9 plateau in the kagome antiferromagnet  $\rm YCu_3(OH)_{6+\it x}Br_{3-\it x}$ ($x\approx0.5$), confirming the stability of this plateau under experimental conditions~\cite{YCOHB1,YCOHB2,YCOHB3,YCOHB4,YCOHB5}.
However, the precise magnetic structure of this phase remains unresolved.
Magnetocaloric effect measurements indicate that this plateau phase has an energy gap~\cite{YCOHB4}, which rules out the possibility of a gapless spin-liquid state.
Furthermore, an anomaly around $T$ = 0.5~K has been detected in specific heat measurements conducted under a magnetic field of 16~T~\cite{YCOHB4}, which is just enough to reach the 1/9 plateau, suggesting a possible phase transition.
This anomaly may indicate a transition to a VBC state with broken symmetry. 

\begin{figure}[tb]
  \centering
  \includegraphics[width=80mm]{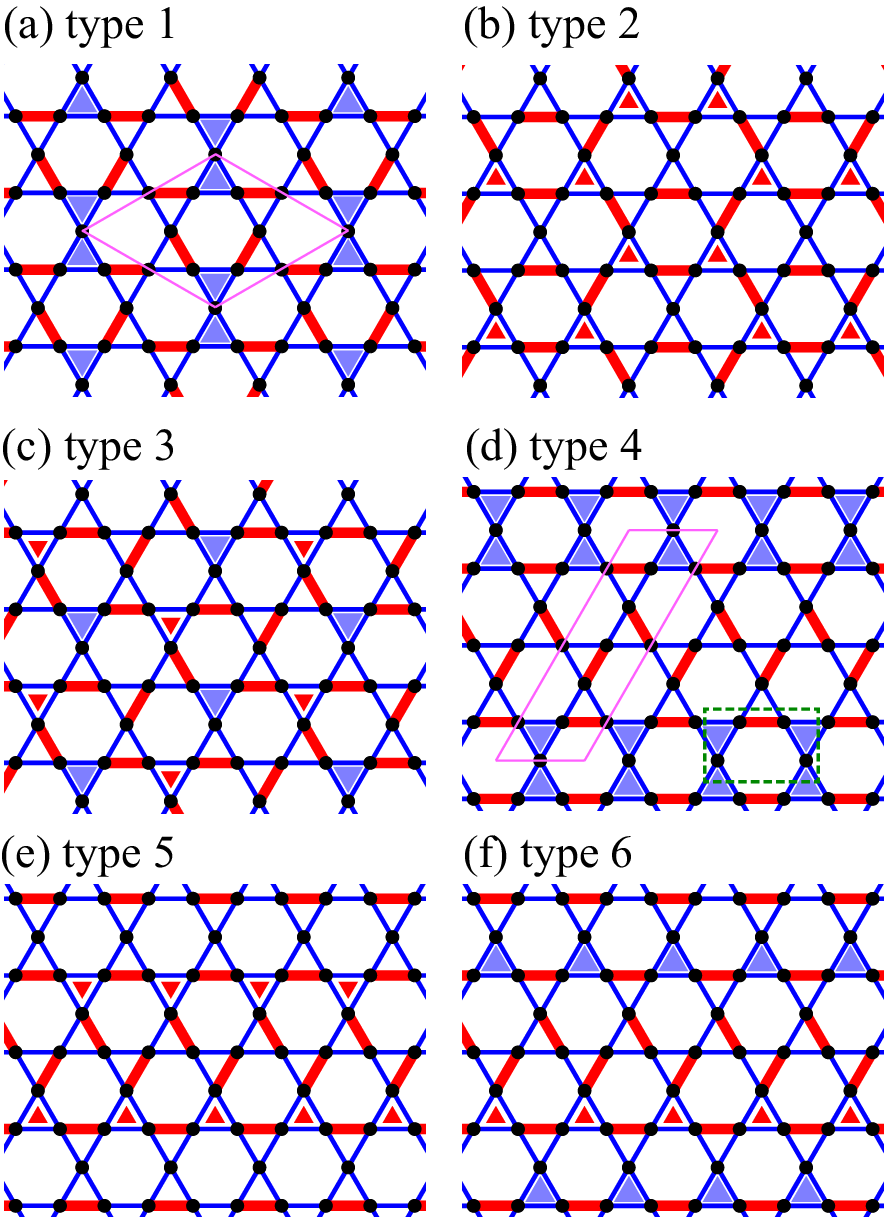}
  \caption {(Color online) Six types of distorted kagome lattices: (a) type~1, (b) type~2, (c) type~3, (d) type~4, (e) type~5, and (f) type~6. Red thick lines and blue lines represent $J_{\rm d}$ and $J$ bonds, respectively, while black circles indicate spin sites. For all types, the ratio of $J$ bonds to $J_{\rm d}$ bonds is $7:2$. Types 1, 2, and 3 feature a unit cell with a $\sqrt{3} \times \sqrt{3}$ structure, shown as the pink rhombus in (a). In contrast, types 4, 5, and 6 exhibit a $1 \times 3$ unit cell, shown as the pink parallelogram in (d). 
The $J$ bonds, indicated by blue and red triangles, contribute to the second-order perturbation process.
Blue triangles indicate regions where a monomer is present and a $J_{\rm d}$ bond is absent, while the red triangles denote regions where neither a monomer nor a $J_{\rm d}$ bond is present.
The green dashed rectangle in (d) indicates the region corresponding to the perturbation process in Fig.~\ref{F4}(c).
  \label{F1}}
\end{figure}

In light of these theoretical and experimental developments, this study focuses on investigating the ground state of the spin-1/2 kagome antiferromagnet at the 1/9 magnetization plateau, particularly from the perspective of VBCs. We aim to find the VBC state with the lowest energy among all possible configurations. According to the Oshikawa-Yamanaka-Affleck criterion~\cite{OYA,OYA2D,OYAKagome}, for the 1/9 plateau to appear in the kagome lattice, the periodicity of the ground state must be a multiple of 9. 
Therefore, we first examine all possible VBC states with $\sqrt{3} \times \sqrt{3}$ and $1 \times 3$ distortions. The VBC states with $\sqrt{3} \times \sqrt{3}$ distortions are limited to the three types  shown in Figs.\ref{F1}(a), \ref{F1}(b), and \ref{F1}(c), while those with $1 \times 3$ distortions are limited to the three types shown in Figs.\ref{F1}(d), \ref{F1}(e), and \ref{F1}(f).
Using the Lanczos-type exact diagonalization method, we calculate the ground-state energies of these six types of VBC distortions on the kagome lattice. The results reveal that type~1 VBC consistently exhibits the lowest ground-state energy, regardless of the distortion strength $J_{\rm d}$.
Analysis of the second derivative of the energy with respect to $J_{\rm d}$ shows that, in the thermodynamic limit, the type~1 VBC state remains stable down to $J_{\rm d} = 1$, corresponding to the undistorted kagome lattice.
 Second-order perturbation theory also supports the robustness of the type~1 VBC state, with energy values agreeing with exact diagonalization results within 3\%. 
Additionally, even when other periodicities are taken into account, it becomes clear that the type~1 VBC still has the lowest energy.
 These findings suggest that the 1/9 magnetization plateau state in the kagome lattice is the type~1 VBC state. 
 Therefore, our results support the findings of previous studies that predicted the VBC ground state~\cite{KLMH5,KLMH8}.
We believe that our findings advance the understanding of magnetization plateau phenomena in frustrated quantum spin systems.

The Hamiltonian for the spin-1/2 kagome lattice with VBC distortions is defined as
\begin{eqnarray} 
H &=& \sum_{\langle i,j \rangle }J_{i,j} \mathbf{S}_i \cdot \mathbf{S}_j,
\end{eqnarray}
where $\mathbf{S}_i$ is the spin-$\frac{1}{2}$ operator at site $i$, $\langle i,j \rangle$ runs over the nearest-neighbor spin pairs, $J_{i,j}$ corresponds to either $J_{\rm d}$ or $J$ shown in Fig.~\ref{F1}.  In the following we set $J =1$ as the energy unit.
We perform Lanczos-type exact diagonalization calculations at zero temperature on distorted kagome lattices, as shown in Fig.~S1, with system sizes of $N = 18, 27,$ and 36, under periodic boundary conditions.

First, we calculate the energy of the kagome lattice with six types of distortions at $M/M_{\rm {sat}} = 1/9$. 
Table~\ref{E27} shows the calculation results for the ground-state energy per site for the $N = 27$ cluster, where type~1 has the lowest energy for all values of $J_{\rm d} = 1.05$, $1.1$, $1.2$, and $1.5$, followed by 
type~3.
Table~\ref{E36} shows the calculation results for the energy for the $N = 36$ cluster. 
In this case, the calculations were performed for type~1, type~2 and type~3, while type~4, type~5, and type~6 were excluded because the cluster shape does not match for these types. Specifically, the green rhombus region shown in Fig.~S1(a) is not applicable to type~4, type~5, and type~6.
The results in Table~\ref{E36} indicate that even for $N = 36$, type~1 has the lowest energy.
Thus, in the undistorted kagome lattice, the type~1 VBC is considered the most promising candidate for the ground state of the 1/9 plateau.

\begin{table}[tb]
\caption{Calculation results for the ground-state energy of the $N = 27$ cluster for six types of distortions.
}
  \begin{tabular}{ c | c c c c } \hline 
              &  \multicolumn{4}{c}{$E/N$} \\ 
              & $J_{\rm d}$=1.05 & 1.1 & 1.2 & 1.5  \\  \hline  
   type~1  & -0.4357261  & -0.4477514  & -0.4743594  & -0.5613539   \\ 
   type~2  & -0.4329215  & -0.4418721  & -0.4651165  & -0.5480765   \\ 
   type~3  & -0.4347572  & -0.4457179  & -0.4706515  & -0.5551643   \\  \hline
   type~4  & -0.4337044  & -0.4441277  & -0.4686113  & -0.5517407   \\ 
   type~5  & -0.4329368  & -0.4420690  & -0.4634804  & -0.5467859   \\ 
   type~6  & -0.4339366  & -0.4441060 & -0.4668058  & -0.5477835    \\  \hline 
  \end{tabular}
\label{E27}
\end{table}
\begin{table}[tb]
\caption{Calculation results for the ground-state energy of the $N = 36$ cluster for three types of distortions.}
  \begin{tabular}{ c | c c c c } \hline 
              &  \multicolumn{4}{c}{$E/N$} \\ 
              & $J_{\rm d}$=1.05 & 1.1 & 1.2 & 1.5  \\  \hline  
   type~1  & -0.4345040  & -0.4469142  & -0.4739212  & -0.5612213   \\ 
   type~2  & -0.4312615  & -0.4403666  & -0.4637512  & -0.5470755   \\ 
   type~3  & -0.4335626  & -0.4449669  & -0.4704103  & -0.5552462   \\  \hline
  \end{tabular}
\label{E36}
\end{table}

\begin{figure}[b]
  \centering
  \includegraphics[width=70mm]{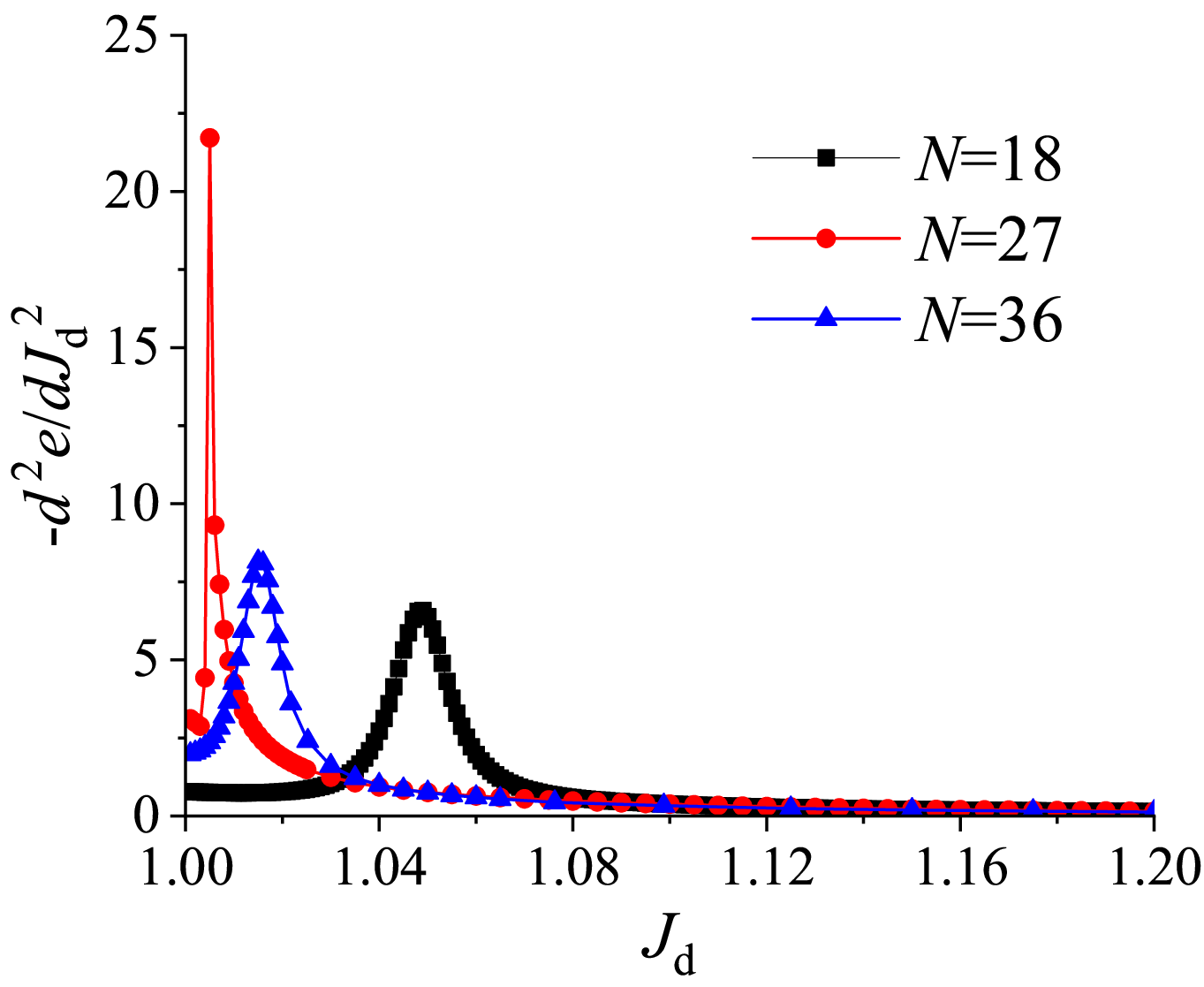}
  \caption{(Color online) Calculation results for the second derivative of the energy density $e=E/N$ with respect to  $J_{\rm d}$.
  \label{F2}}
\end{figure}

Next, we investigate the stability of the type~1 VBC. In the limit of $J_{\rm d} \to \infty$, the ground state forms exact singlet dimers on the $J_{\rm d}$ bonds, while the remaining spins align parallel to the magnetic field, resulting in the emergence of the $1/9$ plateau.
Figure~\ref{F2} shows the calculation results for the second derivative of the energy per site, $e = E/N$, with respect to $J_{\rm d}$.
Peaks that are considered to correspond to phase transitions were observed for each system with $N=18$, $27$, and $36$. These peaks are located at $J_{\rm d}=1.049$ for $N=18$, $J_{\rm d}=1.005$ for $N=27$, and $J_{\rm d}=1.015$ for $N=36$.
Even for the largest system in the present calculations, with $N=36$, it is found that the type~1 VBC state becomes the ground state if a distortion of only 1.5\% is present. 
Furthermore, compared to $N=18$, the peak position for $N=36$ is closer to $J_{\rm d}=1$, suggesting that in the limit $N \to \infty$, the VBC state is expected to remain the ground state down to $J_{\rm d}=1$.
However, since the $J_{\rm d}$ value at the peak position is the smallest for $N = 27$, the peak does not necessarily approach $J_{\rm d} = 1$ as the system size increases. The reason for this behavior is that, at $J_{\rm d} = 1$, the ground state exhibits fourfold degeneracy for $N = 27$, whereas no degeneracy is observed for $N = 18$ and $N = 36$, indicating a unique behavior for $N = 27$.

\begin{figure}[tb]
  \centering
  \includegraphics[width=86mm]{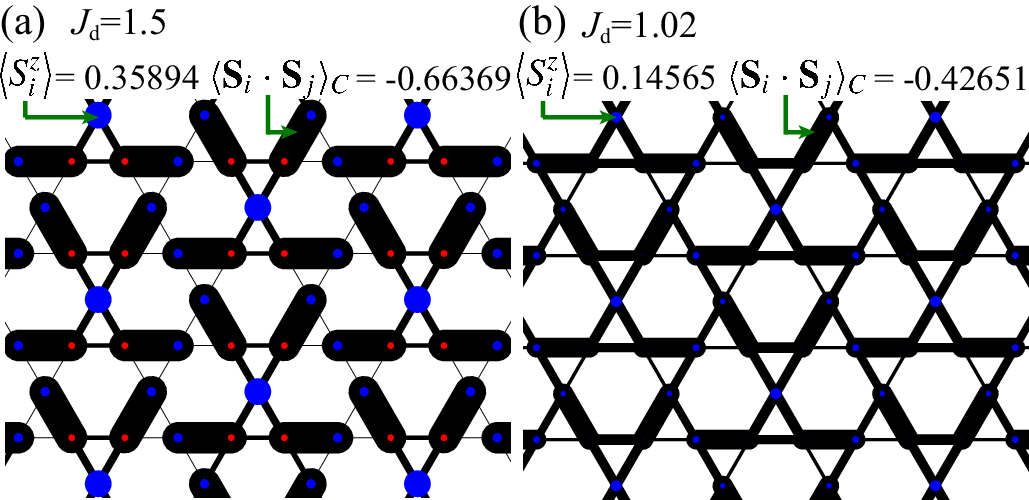}
  \caption{(Color online) Calculation results of the nearest-neighbor spin-spin correlations $\langle \mathbf{S}_i\cdot\mathbf{S}_j\rangle - \langle S_i^z\rangle\langle S_j^z\rangle$ and local magnetizations $\langle S_i^z\rangle$ in the $N = 36$ cluster for type~1. 
(a) $J_{\rm d}=1.5$, (b) $J_{\rm d}=1.02$.  
The width of the black solid lines represents the magnitude of $\langle \mathbf{S}_i\cdot\mathbf{S}_j\rangle - \langle S_i^z\rangle\langle S_j^z\rangle$,which are all negative.
Blue (red) circles on each site denote the positive (negative) value of $\langle S_i^z\rangle$, and their diameter represents its magnitude.
  \label{F3}}
\end{figure}

We investigate the magnetic structure of type~1. 
Figure~\ref{F3} shows the nearest-neighbor spin correlation $\langle  \mathbf{S}_i\cdot\mathbf{S}_j\rangle_C = \langle \mathbf{S}_i\cdot\mathbf{S}_j\rangle - \langle S_i^z\rangle\langle S_j^z\rangle$ and  the local magnetization $\langle S_i^z\rangle$.  The thickness of the lines represents the strength of the correlations. 
Blue (red) circles on each site denote the positive (negative) value of $\langle S_i^z\rangle$, and the diameter of the circles  corresponds to its magnitude.
At $J_{\rm d} = 1.5$, as shown in Fig.~\ref{F3}(a), strong correlations are observed on the $J_{\rm d}$ bonds, indicating that they almost form singlet dimers. The remaining spins form nearly up-spin monomers.
In Fig.~\ref{F3}(b), despite $J_{\rm d}$ being only 2\% larger than $J$, i.e., $J_{\rm d} = 1.02$, the correlations on the $J_{\rm d}$ bonds are the strongest. Additionally, the $\langle S_i^z \rangle$ values for spins not belonging to the $J_{\rm d}$ bonds are the largest. These observations indicate that the VBC state remains stable even at $J_{\rm d} = 1.02$.
As a reference, at $J_{\rm d} = 1$, the values are $\langle \mathbf{S}_i \cdot \mathbf{S}_j \rangle_C = -0.2157770$ and $\langle S_i^z \rangle = 0.0555556$.

\begin{figure}[tb]
  \centering
  \includegraphics[width=65mm]{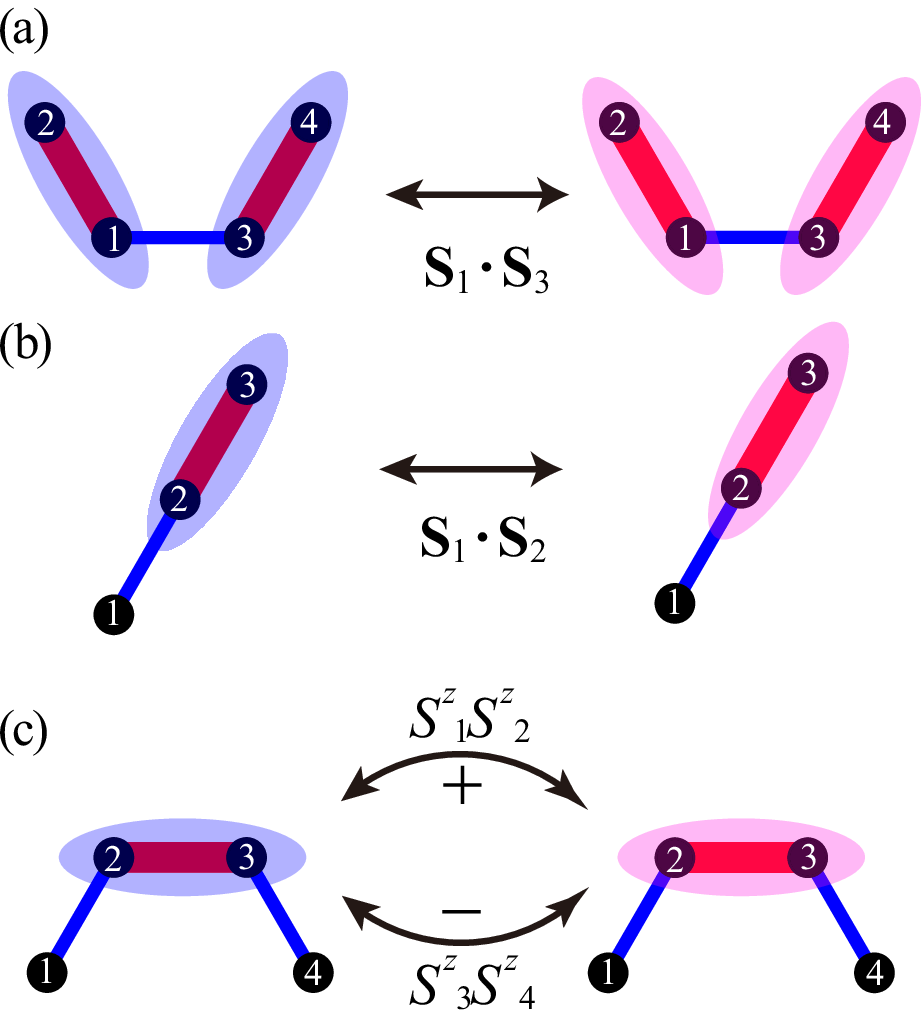}
  \caption{(Color online) Schematic diagrams of the second-order perturbation process. (a) Between dimers. (b) Between a monomer and a dimer. (c) Corresponding to the region marked by the green dashed rectangle in Fig.~\ref{F1}(d).
The blue and red ellipses represent singlet and triplet dimers, respectively.
  \label{F4}}
\end{figure}

 Here, a question arises: as shown in Tables~\ref{E27} and ~\ref{E36}, why does type~1 have the lowest energy among the six types of distortions?
 The reason is clarified below using perturbation theory.
The analysis begins with the case of $J_{\rm d} \gg J (=1)$.
The unperturbed Hamiltonian $H_0$ consists only of the $J_{\rm d}$ bonds, and its ground state forms singlets on all the 
$J_{\rm d}$ bonds, with the energy given by $E_0^{(0)}=-\frac{3}{4} J_{\rm d} \times \frac{4}{9} N$.
 The $J$ terms are treated as a perturbation, denoted by $H'$. It can be readily verified that the first-order perturbation energy vanishes.
  Accordingly, the second-order perturbation is examined in detail.
  The second-order perturbation energy $E_0^{(2)}$ is expressed as follows:
\begin{eqnarray} \label{SeE}
E_0^{(2)} &=& \sum_{ n \neq 0 } \frac{ | \langle n^{(0)} | H' | 0^{(0)} \rangle |^2}{E_0^{(0)} -E_n^{(0)} },
\end{eqnarray}
where $| n^{(0)} \rangle$ is an eigenstate of $H_0$ with the eigenenergy $E_n^{(0)}$.
 First, we consider type~1, type~2, and type~3. The blue triangles and small red triangles shown in Fig.~\ref{F1} contribute to the second-order perturbation for the $J$ bonds. 
 All other $J$ bonds give zero contribution.
This is because the singlet dimer on the $J_{\rm d}$ bond interacts with a single site through two $J$ bonds, where the contributions from each $J$ bond have opposite signs in the second-order perturbation. 
As a result, these contributions cancel out, leading to a net contribution of zero.
The perturbation process between the singlet dimers on the $J_{\rm d}$ bonds is shown in Fig.~\ref{F4}(a). In this process, two singlet dimers become triplets. In contrast, the perturbation process between a singlet dimer and an up-spin monomer is illustrated in Fig.~\ref{F4}(b), where a singlet dimer becomes a triplet. 
In these two processes, the denominator in Eq.~(\ref{SeE}) is found to be twice as large for the perturbation between dimers compared to the perturbation between a dimer and a monomer (note that the numerator has the same value in both cases).
Therefore, the larger the number of the perturbation processes involving the monomers, that is, the structures with more blue triangles shown in Fig.~\ref{F1}, the lower the resulting energy.
This explains the relationship among the ground-state energies shown in Tables~\ref{E27} and~\ref{E36}, namely, type~1 $<$ type~3 $<$ type~2.
However, this reasoning alone cannot explain why the energy of type~4 is higher than that of type~1.
This issue can be resolved by focusing on the green rectangle in Fig.~\ref{F1}(d). 
The perturbation processes for the four spins within this green rectangle are illustrated in Fig.~\ref{F4}(c).
 In this spin configuration, the terms $S_1^zS_2^z$ and $S_3^zS_4^z$ contribute as perturbations, but their contributions cancel out due to opposite signs, reducing the net contribution from perturbations.
Such processes do not exist in type~1, type~2, and type~3, which contributes to the higher energy of type~4.
For reference, Table~\ref{sop} shows the second-order perturbation energy densities ($E_0^{(2)} /N$) for the six types.

We compare the results of exact diagonalization with those from perturbation theory. 
At $J_{\rm d} = 1.5$, the total energy density for type~1, including both zeroth- and second-order perturbations, is   $e_{\rm{pe}} = -0.5694444$. When compared to the exact diagonalization energy for $N = 36$, $e_{\rm{36ED}}$, the relative difference is given by $(e_{\rm{pe}} - e_{\text{36ED}}) / e_{\rm{36ED}} = 0.0147$, which is less than 1.5\%. 
Furthermore, for $J_{\rm d} = 1 (=J)$, the exact diagonalization energy density for $N = 36$ is $e_{\text{36ED}} = -0.4253813$, while the perturbation energy is $e_{\rm{pe}} = -0.4375$, resulting in a relative difference of 0.0285, which is less than 3\%. 
The high accuracy achieved with the second-order perturbation suggests strong evidence that the ground state of the 1/9 plateau in the kagome lattice is the type~1 VBC state. 
Performing additional higher-order perturbation calculations would enable even more accurate energy estimates.

\begin{table}[tb]
\caption{The second-order perturbation energy density for the six types.}
  \begin{tabular}{ c c c c c c } \hline 
   type~1  & type~2   & type~3  & type~4  &  type~5 &  type~6 \\ 
   $-\frac{5J^2}{48J_{\rm d}}$  & $-\frac{3J^2}{48J_{\rm d}}$  & $-\frac{4J^2}{48J_{\rm d}}$  & $-\frac{11J^2}{144J_{\rm d}}$  & $-\frac{3J^2}{48J_{\rm d}}$ &  $-\frac{5J^2}{72J_{\rm d}}$ \\  \hline
  \end{tabular}
\label{sop}
\end{table}

Why does perturbation theory remain a good approximation even at $J_{\rm d} =1$?
 The reason is that when the $J$ belonging to the blue triangles are set to zero, 
 while all other $J$ and $J_{\rm d}$ are set to 1, the type~1 VBC state becomes the exact ground state. Consequently, even at $J_{\rm d}=1$, the $J$-bonds belonging to the blue triangles can still be treated as a perturbation.

Finally, we consider the possibility of VBC states with other periodicities. 
For all dimer-monomer states which are VBC states, the total number of blue and red triangles shown in Fig.~\ref{F1} remains the same. 
Moving the position of one of the blue triangles in type~1 to a neighboring site increases the second-order perturbation energy through the perturbation process shown in Fig~\ref{F4}(c). 
Therefore, the positions of the blue triangles must be equivalent to those in Fig.~\ref{F1}(a).
By considering structures with distortion periods larger than 9, it is possible to create other states with the same arrangement of blue triangles as in type~1.
For example, in the case of clusters with periodic boundary conditions of $N = L\sqrt{3} \times L\sqrt{3} \times 3 = 9L^2$, there are $9 \times 2^L $ structures.
This is due to the existence of $L$ independent $\Delta$-chains arranged along the vertical axis as shown in Fig.~S2 when the $J$ value of the blue triangles shown in Fig.~\ref{F1}(a) is 0 (note that periodic boundary conditions are assumed). 
When $J_{\rm d} = J$, each $\Delta$-chain has a twofold degeneracy, meaning there are two possible arrangements of singlet dimers.
The factor of 9 in $9 \times 2^L$ arises from the choice of which site among the $\sqrt{3} \times \sqrt{3} \times 3$ sites becomes the up-spin monomer.
Here, we consider the case of $L=2$, i.e., $N=36$.
Although there are 36 possible distortions, only two inequivalent groups exist, one of which is type~1.
For the other distortion shown in Fig.~S3, at $J_{\rm d}=1.1$, an energy of $e = -0.4466448$ is obtained, which is higher than the energy of type~1 ($e = -0.4477514$). Therefore, even if long-period VBC is considered as a candidate, we can conclude that type~1 has the lowest energy.
At $J_{\rm d} = 1$, the type~1 VBC state involves the breaking of translational, rotational, and mirror symmetries, leading to an 18-fold degeneracy~\cite{KLMH5}.

In this study, we investigated the ground state of the spin-1/2 kagome antiferromagnet at the 1/9 magnetization plateau, focusing on VBC distortions. Using  the Lanczos-type exact  diagonalization of six types of distorted kagome lattices, we found that the type~1 VBC consistently exhibits the lowest ground-state energy, regardless of the distortion strength $J_{\rm d}$.
Further analysis using the second derivative of the energy with respect to $J_{\rm d}$ revealed that the type~1 VBC remains stable down to $J_{\rm d} = 1$, which corresponds to the undistorted kagome lattice.
Additionally, the energy values obtained from second-order perturbation theory were consistent with the exact diagonalization results within 3\%, further supporting the stability of the type~1 VBC state.
We also explored the possibility that other types of VBC distortions could become the ground state, but type~1 VBC still exhibited the lowest energy.
These findings support the conclusion that the 1/9 magnetization plateau in the kagome lattice is the VBC state, consistent with previous theoretical predictions~\cite{KLMH5,KLMH8}. 
Due to the broken translational symmetry inherent in the type~1 VBC state, a phase transition is expected to occur at low temperatures under a magnetic field sufficient to reach the 1/9 plateau. 
An anomaly suggesting a phase transition has already been observed in specific heat measurements conducted on $\rm YCu_3(OH)_{6+\it x}Br_{3-\it x}$~($x\approx0.5$)~\cite{YCOHB4}.
Therefore, investigating its magnetic structure may reveal the presence of the type~1 VBC.
We hope that future experiments will clarify the magnetic structure of the 1/9 magnetization plateau and allow for a comparison with the results presented in this study.

\begin{acknowledgments}
We thank Y. Fukumoto and S. C Furuya for useful discussions.
We would like to thank the Supercomputer Center, Institute for Solid State Physics, University of Tokyo for the use of their facilities.
\end{acknowledgments} 
\bibliography{ref.bib} 

%apsrev4-2.bst 2019-01-14 (MD) hand-edited version of apsrev4-1.bst
%Control: key (0)
%Control: author (8) initials jnrlst
%Control: editor formatted (1) identically to author
%Control: production of article title (0) allowed
%Control: page (0) single
%Control: year (1) truncated
%Control: production of eprint (0) enabled
\begin{thebibliography}{34}%
\makeatletter
\providecommand \@ifxundefined [1]{%
 \@ifx{#1\undefined}
}%
\providecommand \@ifnum [1]{%
 \ifnum #1\expandafter \@firstoftwo
 \else \expandafter \@secondoftwo
 \fi
}%
\providecommand \@ifx [1]{%
 \ifx #1\expandafter \@firstoftwo
 \else \expandafter \@secondoftwo
 \fi
}%
\providecommand \natexlab [1]{#1}%
\providecommand \enquote  [1]{``#1''}%
\providecommand \bibnamefont  [1]{#1}%
\providecommand \bibfnamefont [1]{#1}%
\providecommand \citenamefont [1]{#1}%
\providecommand \href@noop [0]{\@secondoftwo}%
\providecommand \href [0]{\begingroup \@sanitize@url \@href}%
\providecommand \@href[1]{\@@startlink{#1}\@@href}%
\providecommand \@@href[1]{\endgroup#1\@@endlink}%
\providecommand \@sanitize@url [0]{\catcode `\\12\catcode `\$12\catcode
  `\&12\catcode `\#12\catcode `\^12\catcode `\_12\catcode `\%12\relax}%
\providecommand \@@startlink[1]{}%
\providecommand \@@endlink[0]{}%
\providecommand \url  [0]{\begingroup\@sanitize@url \@url }%
\providecommand \@url [1]{\endgroup\@href {#1}{\urlprefix }}%
\providecommand \urlprefix  [0]{URL }%
\providecommand \Eprint [0]{\href }%
\providecommand \doibase [0]{https://doi.org/}%
\providecommand \selectlanguage [0]{\@gobble}%
\providecommand \bibinfo  [0]{\@secondoftwo}%
\providecommand \bibfield  [0]{\@secondoftwo}%
\providecommand \translation [1]{[#1]}%
\providecommand \BibitemOpen [0]{}%
\providecommand \bibitemStop [0]{}%
\providecommand \bibitemNoStop [0]{.\EOS\space}%
\providecommand \EOS [0]{\spacefactor3000\relax}%
\providecommand \BibitemShut  [1]{\csname bibitem#1\endcsname}%
\let\auto@bib@innerbib\@empty
%</preamble>
\bibitem [{\citenamefont {Yin}\ \emph {et~al.}(2022)\citenamefont {Yin},
  \citenamefont {Lian},\ and\ \citenamefont {Hasan}}]{KLR}%
  \BibitemOpen
  \bibfield  {author} {\bibinfo {author} {\bibfnamefont {J.-X.}\ \bibnamefont
  {Yin}}, \bibinfo {author} {\bibfnamefont {B.}~\bibnamefont {Lian}},\ and\
  \bibinfo {author} {\bibfnamefont {M.~Z.}\ \bibnamefont {Hasan}},\ }\bibfield
  {title} {\bibinfo {title} {Topological kagome magnets and superconductors},\
  }\href {https://doi.org/10.1038/s41586-022-05516-0} {\bibfield  {journal}
  {\bibinfo  {journal} {Nature}\ }\textbf {\bibinfo {volume} {612}},\ \bibinfo
  {pages} {647} (\bibinfo {year} {2022})}\BibitemShut {NoStop}%
\bibitem [{\citenamefont {Yan}\ \emph {et~al.}(2011)\citenamefont {Yan},
  \citenamefont {Huse},\ and\ \citenamefont {White}}]{KLZ2-1}%
  \BibitemOpen
  \bibfield  {author} {\bibinfo {author} {\bibfnamefont {S.}~\bibnamefont
  {Yan}}, \bibinfo {author} {\bibfnamefont {D.~A.}\ \bibnamefont {Huse}},\ and\
  \bibinfo {author} {\bibfnamefont {S.~R.}\ \bibnamefont {White}},\ }\bibfield
  {title} {\bibinfo {title} {Spin-liquid ground state of the s=1/2 kagome
  heisenberg antiferromagnet},\ }\href
  {https://www.science.org/doi/10.1126/science.1201080} {\bibfield  {journal}
  {\bibinfo  {journal} {Science}\ }\textbf {\bibinfo {volume} {332}},\ \bibinfo
  {pages} {1173} (\bibinfo {year} {2011})}\BibitemShut {NoStop}%
\bibitem [{\citenamefont {Depenbrock}\ \emph {et~al.}(2012)\citenamefont
  {Depenbrock}, \citenamefont {McCulloch},\ and\ \citenamefont
  {Schollw\"ock}}]{KLZ2-2}%
  \BibitemOpen
  \bibfield  {author} {\bibinfo {author} {\bibfnamefont {S.}~\bibnamefont
  {Depenbrock}}, \bibinfo {author} {\bibfnamefont {I.~P.}\ \bibnamefont
  {McCulloch}},\ and\ \bibinfo {author} {\bibfnamefont {U.}~\bibnamefont
  {Schollw\"ock}},\ }\bibfield  {title} {\bibinfo {title} {Nature of the
  spin-liquid ground state of the $s=1/2$ heisenberg model on the kagome
  lattice},\ }\href {https://doi.org/10.1103/PhysRevLett.109.067201} {\bibfield
   {journal} {\bibinfo  {journal} {Phys. Rev. Lett.}\ }\textbf {\bibinfo
  {volume} {109}},\ \bibinfo {pages} {067201} (\bibinfo {year}
  {2012})}\BibitemShut {NoStop}%
\bibitem [{\citenamefont {Mei}\ \emph {et~al.}(2017)\citenamefont {Mei},
  \citenamefont {Chen}, \citenamefont {He},\ and\ \citenamefont
  {Wen}}]{KLZ2-3}%
  \BibitemOpen
  \bibfield  {author} {\bibinfo {author} {\bibfnamefont {J.-W.}\ \bibnamefont
  {Mei}}, \bibinfo {author} {\bibfnamefont {J.-Y.}\ \bibnamefont {Chen}},
  \bibinfo {author} {\bibfnamefont {H.}~\bibnamefont {He}},\ and\ \bibinfo
  {author} {\bibfnamefont {X.-G.}\ \bibnamefont {Wen}},\ }\bibfield  {title}
  {\bibinfo {title} {Gapped spin liquid with $z_2$ topological order for the
  kagome heisenberg model},\ }\href
  {https://doi.org/10.1103/PhysRevB.95.235107} {\bibfield  {journal} {\bibinfo
  {journal} {Phys. Rev. B}\ }\textbf {\bibinfo {volume} {95}},\ \bibinfo
  {pages} {235107} (\bibinfo {year} {2017})}\BibitemShut {NoStop}%
\bibitem [{\citenamefont {Ran}\ \emph {et~al.}(2007)\citenamefont {Ran},
  \citenamefont {Hermele}, \citenamefont {Lee},\ and\ \citenamefont
  {Wen}}]{KLU1-1}%
  \BibitemOpen
  \bibfield  {author} {\bibinfo {author} {\bibfnamefont {Y.}~\bibnamefont
  {Ran}}, \bibinfo {author} {\bibfnamefont {M.}~\bibnamefont {Hermele}},
  \bibinfo {author} {\bibfnamefont {P.~A.}\ \bibnamefont {Lee}},\ and\ \bibinfo
  {author} {\bibfnamefont {X.-G.}\ \bibnamefont {Wen}},\ }\bibfield  {title}
  {\bibinfo {title} {Projected-wave-function study of the spin-$1/2$ heisenberg
  model on the kagom\'e lattice},\ }\href
  {https://doi.org/10.1103/PhysRevLett.98.117205} {\bibfield  {journal}
  {\bibinfo  {journal} {Phys. Rev. Lett.}\ }\textbf {\bibinfo {volume} {98}},\
  \bibinfo {pages} {117205} (\bibinfo {year} {2007})}\BibitemShut {NoStop}%
\bibitem [{\citenamefont {Iqbal}\ \emph {et~al.}(2011)\citenamefont {Iqbal},
  \citenamefont {Becca},\ and\ \citenamefont {Poilblanc}}]{KLU1-2}%
  \BibitemOpen
  \bibfield  {author} {\bibinfo {author} {\bibfnamefont {Y.}~\bibnamefont
  {Iqbal}}, \bibinfo {author} {\bibfnamefont {F.}~\bibnamefont {Becca}},\ and\
  \bibinfo {author} {\bibfnamefont {D.}~\bibnamefont {Poilblanc}},\ }\bibfield
  {title} {\bibinfo {title} {Valence-bond crystal in the extended kagome
  spin-$\frac{1}{2}$ quantum heisenberg antiferromagnet: A variational monte
  carlo approach},\ }\href {https://doi.org/10.1103/PhysRevB.83.100404}
  {\bibfield  {journal} {\bibinfo  {journal} {Phys. Rev. B}\ }\textbf {\bibinfo
  {volume} {83}},\ \bibinfo {pages} {100404(R)} (\bibinfo {year}
  {2011})}\BibitemShut {NoStop}%
\bibitem [{\citenamefont {Iqbal}\ \emph {et~al.}(2013)\citenamefont {Iqbal},
  \citenamefont {Becca}, \citenamefont {Sorella},\ and\ \citenamefont
  {Poilblanc}}]{KLU1-3}%
  \BibitemOpen
  \bibfield  {author} {\bibinfo {author} {\bibfnamefont {Y.}~\bibnamefont
  {Iqbal}}, \bibinfo {author} {\bibfnamefont {F.}~\bibnamefont {Becca}},
  \bibinfo {author} {\bibfnamefont {S.}~\bibnamefont {Sorella}},\ and\ \bibinfo
  {author} {\bibfnamefont {D.}~\bibnamefont {Poilblanc}},\ }\bibfield  {title}
  {\bibinfo {title} {Gapless spin-liquid phase in the kagome spin-$\frac{1}{2}$
  heisenberg antiferromagnet},\ }\href
  {https://doi.org/10.1103/PhysRevB.87.060405} {\bibfield  {journal} {\bibinfo
  {journal} {Phys. Rev. B}\ }\textbf {\bibinfo {volume} {87}},\ \bibinfo
  {pages} {060405(R)} (\bibinfo {year} {2013})}\BibitemShut {NoStop}%
\bibitem [{\citenamefont {Liao}\ \emph {et~al.}(2017)\citenamefont {Liao},
  \citenamefont {Xie}, \citenamefont {Chen}, \citenamefont {Liu}, \citenamefont
  {Xie}, \citenamefont {Huang}, \citenamefont {Normand},\ and\ \citenamefont
  {Xiang}}]{KLU1-4}%
  \BibitemOpen
  \bibfield  {author} {\bibinfo {author} {\bibfnamefont {H.~J.}\ \bibnamefont
  {Liao}}, \bibinfo {author} {\bibfnamefont {Z.~Y.}\ \bibnamefont {Xie}},
  \bibinfo {author} {\bibfnamefont {J.}~\bibnamefont {Chen}}, \bibinfo {author}
  {\bibfnamefont {Z.~Y.}\ \bibnamefont {Liu}}, \bibinfo {author} {\bibfnamefont
  {H.~D.}\ \bibnamefont {Xie}}, \bibinfo {author} {\bibfnamefont {R.~Z.}\
  \bibnamefont {Huang}}, \bibinfo {author} {\bibfnamefont {B.}~\bibnamefont
  {Normand}},\ and\ \bibinfo {author} {\bibfnamefont {T.}~\bibnamefont
  {Xiang}},\ }\bibfield  {title} {\bibinfo {title} {Gapless spin-liquid ground
  state in the $s=1/2$ kagome antiferromagnet},\ }\href
  {https://doi.org/10.1103/PhysRevLett.118.137202} {\bibfield  {journal}
  {\bibinfo  {journal} {Phys. Rev. Lett.}\ }\textbf {\bibinfo {volume} {118}},\
  \bibinfo {pages} {137202} (\bibinfo {year} {2017})}\BibitemShut {NoStop}%
\bibitem [{\citenamefont {He}\ \emph {et~al.}(2017)\citenamefont {He},
  \citenamefont {Zaletel}, \citenamefont {Oshikawa},\ and\ \citenamefont
  {Pollmann}}]{KLU1-5}%
  \BibitemOpen
  \bibfield  {author} {\bibinfo {author} {\bibfnamefont {Y.-C.}\ \bibnamefont
  {He}}, \bibinfo {author} {\bibfnamefont {M.~P.}\ \bibnamefont {Zaletel}},
  \bibinfo {author} {\bibfnamefont {M.}~\bibnamefont {Oshikawa}},\ and\
  \bibinfo {author} {\bibfnamefont {F.}~\bibnamefont {Pollmann}},\ }\bibfield
  {title} {\bibinfo {title} {Signatures of dirac cones in a dmrg study of the
  kagome heisenberg model},\ }\href {https://doi.org/10.1103/PhysRevX.7.031020}
  {\bibfield  {journal} {\bibinfo  {journal} {Phys. Rev. X}\ }\textbf {\bibinfo
  {volume} {7}},\ \bibinfo {pages} {031020} (\bibinfo {year}
  {2017})}\BibitemShut {NoStop}%
\bibitem [{\citenamefont {Syromyatnikov}\ and\ \citenamefont
  {Maleyev}(2002)}]{KLVBC1}%
  \BibitemOpen
  \bibfield  {author} {\bibinfo {author} {\bibfnamefont {A.~V.}\ \bibnamefont
  {Syromyatnikov}}\ and\ \bibinfo {author} {\bibfnamefont {S.~V.}\ \bibnamefont
  {Maleyev}},\ }\bibfield  {title} {\bibinfo {title} {Hidden long-range order
  in kagom\'e heisenberg antiferromagnets},\ }\href
  {https://doi.org/10.1103/PhysRevB.66.132408} {\bibfield  {journal} {\bibinfo
  {journal} {Phys. Rev. B}\ }\textbf {\bibinfo {volume} {66}},\ \bibinfo
  {pages} {132408} (\bibinfo {year} {2002})}\BibitemShut {NoStop}%
\bibitem [{\citenamefont {Singh}\ and\ \citenamefont {Huse}(2007)}]{KLVBC2}%
  \BibitemOpen
  \bibfield  {author} {\bibinfo {author} {\bibfnamefont {R.~R.~P.}\
  \bibnamefont {Singh}}\ and\ \bibinfo {author} {\bibfnamefont {D.~A.}\
  \bibnamefont {Huse}},\ }\bibfield  {title} {\bibinfo {title} {Ground state of
  the spin-1/2 kagome-lattice heisenberg antiferromagnet},\ }\href
  {https://doi.org/10.1103/PhysRevB.76.180407} {\bibfield  {journal} {\bibinfo
  {journal} {Phys. Rev. B}\ }\textbf {\bibinfo {volume} {76}},\ \bibinfo
  {pages} {180407(R)} (\bibinfo {year} {2007})}\BibitemShut {NoStop}%
\bibitem [{\citenamefont {Hwang}\ \emph {et~al.}(2011)\citenamefont {Hwang},
  \citenamefont {Kim}, \citenamefont {Yu},\ and\ \citenamefont
  {Park}}]{KLVBC3}%
  \BibitemOpen
  \bibfield  {author} {\bibinfo {author} {\bibfnamefont {K.}~\bibnamefont
  {Hwang}}, \bibinfo {author} {\bibfnamefont {Y.~B.}\ \bibnamefont {Kim}},
  \bibinfo {author} {\bibfnamefont {J.}~\bibnamefont {Yu}},\ and\ \bibinfo
  {author} {\bibfnamefont {K.}~\bibnamefont {Park}},\ }\bibfield  {title}
  {\bibinfo {title} {Spin cluster operator theory for the kagome lattice
  antiferromagnet},\ }\href {https://doi.org/10.1103/PhysRevB.84.205133}
  {\bibfield  {journal} {\bibinfo  {journal} {Phys. Rev. B}\ }\textbf {\bibinfo
  {volume} {84}},\ \bibinfo {pages} {205133} (\bibinfo {year}
  {2011})}\BibitemShut {NoStop}%
\bibitem [{\citenamefont {Honecker}\ \emph {et~al.}(2004)\citenamefont
  {Honecker}, \citenamefont {Schulenburg},\ and\ \citenamefont
  {Richter}}]{KLMH1}%
  \BibitemOpen
  \bibfield  {author} {\bibinfo {author} {\bibfnamefont {A.}~\bibnamefont
  {Honecker}}, \bibinfo {author} {\bibfnamefont {J.}~\bibnamefont
  {Schulenburg}},\ and\ \bibinfo {author} {\bibfnamefont {J.}~\bibnamefont
  {Richter}},\ }\bibfield  {title} {\bibinfo {title} {Magnetization plateaus in
  frustrated antiferromagnetic quantum spin models},\ }\href
  {https://doi.org/10.1088/0953-8984/16/11/025} {\bibfield  {journal} {\bibinfo
   {journal} {J. Phys.: Condens. Matter}\ }\textbf {\bibinfo {volume} {16}},\
  \bibinfo {pages} {S749} (\bibinfo {year} {2004})}\BibitemShut {NoStop}%
\bibitem [{\citenamefont {Capponi}\ \emph {et~al.}(2013)\citenamefont
  {Capponi}, \citenamefont {Derzhko}, \citenamefont {Honecker}, \citenamefont
  {L\"auchli},\ and\ \citenamefont {Richter}}]{KLMH3}%
  \BibitemOpen
  \bibfield  {author} {\bibinfo {author} {\bibfnamefont {S.}~\bibnamefont
  {Capponi}}, \bibinfo {author} {\bibfnamefont {O.}~\bibnamefont {Derzhko}},
  \bibinfo {author} {\bibfnamefont {A.}~\bibnamefont {Honecker}}, \bibinfo
  {author} {\bibfnamefont {A.~M.}\ \bibnamefont {L\"auchli}},\ and\ \bibinfo
  {author} {\bibfnamefont {J.}~\bibnamefont {Richter}},\ }\bibfield  {title}
  {\bibinfo {title} {Numerical study of magnetization plateaus in the
  spin-$\frac{1}{2}$ kagome heisenberg antiferromagnet},\ }\href
  {https://doi.org/10.1103/PhysRevB.88.144416} {\bibfield  {journal} {\bibinfo
  {journal} {Phys. Rev. B}\ }\textbf {\bibinfo {volume} {88}},\ \bibinfo
  {pages} {144416} (\bibinfo {year} {2013})}\BibitemShut {NoStop}%
\bibitem [{\citenamefont {Nishimoto}\ \emph {et~al.}(2013)\citenamefont
  {Nishimoto}, \citenamefont {Shibata},\ and\ \citenamefont {Hotta}}]{KLMH4}%
  \BibitemOpen
  \bibfield  {author} {\bibinfo {author} {\bibfnamefont {S.}~\bibnamefont
  {Nishimoto}}, \bibinfo {author} {\bibfnamefont {N.}~\bibnamefont {Shibata}},\
  and\ \bibinfo {author} {\bibfnamefont {C.}~\bibnamefont {Hotta}},\ }\bibfield
   {title} {\bibinfo {title} {Controlling frustrated liquids and solids with an
  applied field in a kagome heisenberg antiferromagnet},\ }\href
  {https://doi.org/10.1038/ncomms3287} {\bibfield  {journal} {\bibinfo
  {journal} {Nat. Commun.}\ }\textbf {\bibinfo {volume} {4}},\ \bibinfo {pages}
  {2287} (\bibinfo {year} {2013})}\BibitemShut {NoStop}%
\bibitem [{\citenamefont {Picot}\ \emph {et~al.}(2016)\citenamefont {Picot},
  \citenamefont {Ziegler}, \citenamefont {Or\'us},\ and\ \citenamefont
  {Poilblanc}}]{KLMH5}%
  \BibitemOpen
  \bibfield  {author} {\bibinfo {author} {\bibfnamefont {T.}~\bibnamefont
  {Picot}}, \bibinfo {author} {\bibfnamefont {M.}~\bibnamefont {Ziegler}},
  \bibinfo {author} {\bibfnamefont {R.}~\bibnamefont {Or\'us}},\ and\ \bibinfo
  {author} {\bibfnamefont {D.}~\bibnamefont {Poilblanc}},\ }\bibfield  {title}
  {\bibinfo {title} {Spin-$s$ kagome quantum antiferromagnets in a field with
  tensor networks},\ }\href {https://doi.org/10.1103/PhysRevB.93.060407}
  {\bibfield  {journal} {\bibinfo  {journal} {Phys. Rev. B}\ }\textbf {\bibinfo
  {volume} {93}},\ \bibinfo {pages} {060407} (\bibinfo {year}
  {2016})}\BibitemShut {NoStop}%
\bibitem [{\citenamefont {Schnack}\ \emph {et~al.}(2020)\citenamefont
  {Schnack}, \citenamefont {Schulenburg}, \citenamefont {Honecker},\ and\
  \citenamefont {Richter}}]{KLMH7}%
  \BibitemOpen
  \bibfield  {author} {\bibinfo {author} {\bibfnamefont {J.}~\bibnamefont
  {Schnack}}, \bibinfo {author} {\bibfnamefont {J.}~\bibnamefont
  {Schulenburg}}, \bibinfo {author} {\bibfnamefont {A.}~\bibnamefont
  {Honecker}},\ and\ \bibinfo {author} {\bibfnamefont {J.}~\bibnamefont
  {Richter}},\ }\bibfield  {title} {\bibinfo {title} {Magnon crystallization in
  the kagome lattice antiferromagnet},\ }\href
  {https://doi.org/10.1103/PhysRevLett.125.117207} {\bibfield  {journal}
  {\bibinfo  {journal} {Phys. Rev. Lett.}\ }\textbf {\bibinfo {volume} {125}},\
  \bibinfo {pages} {117207} (\bibinfo {year} {2020})}\BibitemShut {NoStop}%
\bibitem [{\citenamefont {Fang}\ \emph {et~al.}(2023)\citenamefont {Fang},
  \citenamefont {Xi}, \citenamefont {Ran},\ and\ \citenamefont {Su}}]{KLMH8}%
  \BibitemOpen
  \bibfield  {author} {\bibinfo {author} {\bibfnamefont {D.-z.}\ \bibnamefont
  {Fang}}, \bibinfo {author} {\bibfnamefont {N.}~\bibnamefont {Xi}}, \bibinfo
  {author} {\bibfnamefont {S.-J.}\ \bibnamefont {Ran}},\ and\ \bibinfo {author}
  {\bibfnamefont {G.}~\bibnamefont {Su}},\ }\bibfield  {title} {\bibinfo
  {title} {Nature of the 1/9-magnetization plateau in the spin-$\frac{1}{2}$
  kagome heisenberg antiferromagnet},\ }\href
  {https://doi.org/10.1103/PhysRevB.107.L220401} {\bibfield  {journal}
  {\bibinfo  {journal} {Phys. Rev. B}\ }\textbf {\bibinfo {volume} {107}},\
  \bibinfo {pages} {L220401} (\bibinfo {year} {2023})}\BibitemShut {NoStop}%
\bibitem [{\citenamefont {Chern}(2024)}]{KLMH9}%
  \BibitemOpen
  \bibfield  {author} {\bibinfo {author} {\bibfnamefont {G.-W.}\ \bibnamefont
  {Chern}},\ }\bibfield  {title} {\bibinfo {title} {A kagome antiferromagnet
  reaches its quantum plateau},\ }\href
  {https://doi.org/10.1038/s41567-023-02383-y} {\bibfield  {journal} {\bibinfo
  {journal} {Nat. Phys.}\ }\textbf {\bibinfo {volume} {20}},\ \bibinfo {pages}
  {353} (\bibinfo {year} {2024})}\BibitemShut {NoStop}%
\bibitem [{\citenamefont {Bodaiji}\ \emph {et~al.}(2024)\citenamefont
  {Bodaiji}, \citenamefont {Morita},\ and\ \citenamefont {Fukumoto}}]{YCOHC7}%
  \BibitemOpen
  \bibfield  {author} {\bibinfo {author} {\bibfnamefont {K.}~\bibnamefont
  {Bodaiji}}, \bibinfo {author} {\bibfnamefont {K.}~\bibnamefont {Morita}},\
  and\ \bibinfo {author} {\bibfnamefont {Y.}~\bibnamefont {Fukumoto}},\
  }\bibfield  {title} {\bibinfo {title} {Six magnetization plateau phases in a
  spin-$\frac{1}{2}$ distorted kagome antiferromagnet: Application to
  ${\mathrm{y}}_{3}{\mathrm{cu}}_{9}{(\mathrm{OH})}_{19}{\mathrm{cl}}_{8}$},\
  }\href {https://doi.org/10.1103/PhysRevB.110.104431} {\bibfield  {journal}
  {\bibinfo  {journal} {Phys. Rev. B}\ }\textbf {\bibinfo {volume} {110}},\
  \bibinfo {pages} {104431} (\bibinfo {year} {2024})}\BibitemShut {NoStop}%
\bibitem [{\citenamefont {Nakano}\ and\ \citenamefont {Sakai}(2010)}]{KLMH2}%
  \BibitemOpen
  \bibfield  {author} {\bibinfo {author} {\bibfnamefont {H.}~\bibnamefont
  {Nakano}}\ and\ \bibinfo {author} {\bibfnamefont {T.}~\bibnamefont {Sakai}},\
  }\bibfield  {title} {\bibinfo {title} {Magnetization process of
  kagome-lattice heisenberg antiferromagnet},\ }\href
  {https://doi.org/10.1143/JPSJ.79.053707} {\bibfield  {journal} {\bibinfo
  {journal} {J. Phys. Soc. Jpn.}\ }\textbf {\bibinfo {volume} {79}},\ \bibinfo
  {pages} {053707} (\bibinfo {year} {2010})}\BibitemShut {NoStop}%
\bibitem [{\citenamefont {Nakano}\ and\ \citenamefont {Sakai}(2018)}]{KLMH6}%
  \BibitemOpen
  \bibfield  {author} {\bibinfo {author} {\bibfnamefont {H.}~\bibnamefont
  {Nakano}}\ and\ \bibinfo {author} {\bibfnamefont {T.}~\bibnamefont {Sakai}},\
  }\bibfield  {title} {\bibinfo {title} {Numerical-diagonalization study of
  magnetization process of frustrated spin-1/2 heisenberg antiferromagnets in
  two dimensions: Triangular and kagome lattice antiferromagnets},\ }\href
  {https://doi.org/10.7566/JPSJ.87.063706} {\bibfield  {journal} {\bibinfo
  {journal} {J. Phys. Soc. Jpn.}\ }\textbf {\bibinfo {volume} {87}},\ \bibinfo
  {pages} {063706} (\bibinfo {year} {2018})}\BibitemShut {NoStop}%
\bibitem [{\citenamefont {Schnack}\ \emph {et~al.}(2018)\citenamefont
  {Schnack}, \citenamefont {Schulenburg},\ and\ \citenamefont
  {Richter}}]{Ameltkagome1}%
  \BibitemOpen
  \bibfield  {author} {\bibinfo {author} {\bibfnamefont {J.}~\bibnamefont
  {Schnack}}, \bibinfo {author} {\bibfnamefont {J.}~\bibnamefont
  {Schulenburg}},\ and\ \bibinfo {author} {\bibfnamefont {J.}~\bibnamefont
  {Richter}},\ }\bibfield  {title} {\bibinfo {title} {Magnetism of the $n=42$
  kagome lattice antiferromagnet},\ }\href
  {https://doi.org/10.1103/PhysRevB.98.094423} {\bibfield  {journal} {\bibinfo
  {journal} {Phys. Rev. B}\ }\textbf {\bibinfo {volume} {98}},\ \bibinfo
  {pages} {094423} (\bibinfo {year} {2018})}\BibitemShut {NoStop}%
\bibitem [{\citenamefont {Misawa}\ \emph {et~al.}(2020)\citenamefont {Misawa},
  \citenamefont {Motoyama},\ and\ \citenamefont {Yamaji}}]{Ameltkagome2}%
  \BibitemOpen
  \bibfield  {author} {\bibinfo {author} {\bibfnamefont {T.}~\bibnamefont
  {Misawa}}, \bibinfo {author} {\bibfnamefont {Y.}~\bibnamefont {Motoyama}},\
  and\ \bibinfo {author} {\bibfnamefont {Y.}~\bibnamefont {Yamaji}},\
  }\bibfield  {title} {\bibinfo {title} {Asymmetric melting of a one-third
  plateau in kagome quantum antiferromagnets},\ }\href
  {https://doi.org/10.1103/PhysRevB.102.094419} {\bibfield  {journal} {\bibinfo
   {journal} {Phys. Rev. B}\ }\textbf {\bibinfo {volume} {102}},\ \bibinfo
  {pages} {094419} (\bibinfo {year} {2020})}\BibitemShut {NoStop}%
\bibitem [{\citenamefont {Morita}(2023)}]{Ameltkagome3}%
  \BibitemOpen
  \bibfield  {author} {\bibinfo {author} {\bibfnamefont {K.}~\bibnamefont
  {Morita}},\ }\bibfield  {title} {\bibinfo {title} {Stability of the
  $\frac{1}{3}$ magnetization plateau of the ${J}_{1}\ensuremath{-}{J}_{2}$
  kagome heisenberg model},\ }\href
  {https://doi.org/10.1103/PhysRevB.108.184405} {\bibfield  {journal} {\bibinfo
   {journal} {Phys. Rev. B}\ }\textbf {\bibinfo {volume} {108}},\ \bibinfo
  {pages} {184405} (\bibinfo {year} {2023})}\BibitemShut {NoStop}%
\bibitem [{\citenamefont {Jeon}\ \emph {et~al.}(2024)\citenamefont {Jeon},
  \citenamefont {Wulferding}, \citenamefont {Choi}, \citenamefont {Lee},
  \citenamefont {Nam}, \citenamefont {Kim}, \citenamefont {Lee}, \citenamefont
  {Jang}, \citenamefont {Park}, \citenamefont {Lee}, \citenamefont {Choi},
  \citenamefont {Lee}, \citenamefont {Nojiri},\ and\ \citenamefont
  {Choi}}]{YCOHB1}%
  \BibitemOpen
  \bibfield  {author} {\bibinfo {author} {\bibfnamefont {S.}~\bibnamefont
  {Jeon}}, \bibinfo {author} {\bibfnamefont {D.}~\bibnamefont {Wulferding}},
  \bibinfo {author} {\bibfnamefont {Y.}~\bibnamefont {Choi}}, \bibinfo {author}
  {\bibfnamefont {S.}~\bibnamefont {Lee}}, \bibinfo {author} {\bibfnamefont
  {K.}~\bibnamefont {Nam}}, \bibinfo {author} {\bibfnamefont {K.}~\bibnamefont
  {Kim}}, \bibinfo {author} {\bibfnamefont {M.}~\bibnamefont {Lee}}, \bibinfo
  {author} {\bibfnamefont {T.-H.}\ \bibnamefont {Jang}}, \bibinfo {author}
  {\bibfnamefont {J.-H.}\ \bibnamefont {Park}}, \bibinfo {author}
  {\bibfnamefont {S.}~\bibnamefont {Lee}}, \bibinfo {author} {\bibfnamefont
  {S.}~\bibnamefont {Choi}}, \bibinfo {author} {\bibfnamefont {C.}~\bibnamefont
  {Lee}}, \bibinfo {author} {\bibfnamefont {H.}~\bibnamefont {Nojiri}},\ and\
  \bibinfo {author} {\bibfnamefont {K.}~\bibnamefont {Choi}},\ }\bibfield
  {title} {\bibinfo {title} {One-ninth magnetization plateau stabilized by spin
  entanglement in a kagome antiferromagnet},\ }\href
  {https://doi.org/10.1038/s41567-023-02318-7} {\bibfield  {journal} {\bibinfo
  {journal} {Nat. Phys.}\ }\textbf {\bibinfo {volume} {20}},\ \bibinfo {pages}
  {435} (\bibinfo {year} {2024})}\BibitemShut {NoStop}%
\bibitem [{\citenamefont {Suetsugu}\ \emph
  {et~al.}(2024{\natexlab{a}})\citenamefont {Suetsugu}, \citenamefont {Asaba},
  \citenamefont {Kasahara}, \citenamefont {Kohsaka}, \citenamefont {Totsuka},
  \citenamefont {Li}, \citenamefont {Zhao}, \citenamefont {Li}, \citenamefont
  {Tokunaga},\ and\ \citenamefont {Matsuda}}]{YCOHB2}%
  \BibitemOpen
  \bibfield  {author} {\bibinfo {author} {\bibfnamefont {S.}~\bibnamefont
  {Suetsugu}}, \bibinfo {author} {\bibfnamefont {T.}~\bibnamefont {Asaba}},
  \bibinfo {author} {\bibfnamefont {Y.}~\bibnamefont {Kasahara}}, \bibinfo
  {author} {\bibfnamefont {Y.}~\bibnamefont {Kohsaka}}, \bibinfo {author}
  {\bibfnamefont {K.}~\bibnamefont {Totsuka}}, \bibinfo {author} {\bibfnamefont
  {B.}~\bibnamefont {Li}}, \bibinfo {author} {\bibfnamefont {Y.}~\bibnamefont
  {Zhao}}, \bibinfo {author} {\bibfnamefont {Y.}~\bibnamefont {Li}}, \bibinfo
  {author} {\bibfnamefont {M.}~\bibnamefont {Tokunaga}},\ and\ \bibinfo
  {author} {\bibfnamefont {Y.}~\bibnamefont {Matsuda}},\ }\bibfield  {title}
  {\bibinfo {title} {Emergent spin-gapped magnetization plateaus in a
  spin-$1/2$ perfect kagome antiferromagnet},\ }\href
  {https://doi.org/10.1103/PhysRevLett.132.226701} {\bibfield  {journal}
  {\bibinfo  {journal} {Phys. Rev. Lett.}\ }\textbf {\bibinfo {volume} {132}},\
  \bibinfo {pages} {226701} (\bibinfo {year} {2024}{\natexlab{a}})}\BibitemShut
  {NoStop}%
\bibitem [{\citenamefont {Zheng}\ \emph {et~al.}(2023)\citenamefont {Zheng},
  \citenamefont {Zhu}, \citenamefont {Chen}, \citenamefont {Kang},
  \citenamefont {Zhang}, \citenamefont {Jenkins}, \citenamefont {Chan},
  \citenamefont {Zeng}, \citenamefont {Xu}, \citenamefont {Valenzuela},
  \citenamefont {Blawat}, \citenamefont {Singleton}, \citenamefont {Lee},
  \citenamefont {Li},\ and\ \citenamefont {Li}}]{YCOHB3}%
  \BibitemOpen
  \bibfield  {author} {\bibinfo {author} {\bibfnamefont {G.}~\bibnamefont
  {Zheng}}, \bibinfo {author} {\bibfnamefont {Y.}~\bibnamefont {Zhu}}, \bibinfo
  {author} {\bibfnamefont {K.-W.}\ \bibnamefont {Chen}}, \bibinfo {author}
  {\bibfnamefont {B.}~\bibnamefont {Kang}}, \bibinfo {author} {\bibfnamefont
  {D.}~\bibnamefont {Zhang}}, \bibinfo {author} {\bibfnamefont
  {K.}~\bibnamefont {Jenkins}}, \bibinfo {author} {\bibfnamefont
  {A.}~\bibnamefont {Chan}}, \bibinfo {author} {\bibfnamefont {Z.}~\bibnamefont
  {Zeng}}, \bibinfo {author} {\bibfnamefont {A.}~\bibnamefont {Xu}}, \bibinfo
  {author} {\bibfnamefont {O.~A.}\ \bibnamefont {Valenzuela}}, \bibinfo
  {author} {\bibfnamefont {J.}~\bibnamefont {Blawat}}, \bibinfo {author}
  {\bibfnamefont {J.}~\bibnamefont {Singleton}}, \bibinfo {author}
  {\bibfnamefont {P.~A.}\ \bibnamefont {Lee}}, \bibinfo {author} {\bibfnamefont
  {S.}~\bibnamefont {Li}},\ and\ \bibinfo {author} {\bibfnamefont
  {L.}~\bibnamefont {Li}},\ }\href@noop {} {\bibinfo {title} {Unconventional
  magnetic oscillations in kagome mott insulators}} (\bibinfo {year} {2023}),\
  \Eprint {https://arxiv.org/abs/2310.07989} {arXiv:2310.07989} \BibitemShut
  {NoStop}%
\bibitem [{\citenamefont {Suetsugu}\ \emph
  {et~al.}(2024{\natexlab{b}})\citenamefont {Suetsugu}, \citenamefont {Asaba},
  \citenamefont {Ikemori}, \citenamefont {Sekino}, \citenamefont {Kasahara},
  \citenamefont {Totsuka}, \citenamefont {Li}, \citenamefont {Zhao},
  \citenamefont {Li}, \citenamefont {Kohama},\ and\ \citenamefont
  {Matsuda}}]{YCOHB4}%
  \BibitemOpen
  \bibfield  {author} {\bibinfo {author} {\bibfnamefont {S.}~\bibnamefont
  {Suetsugu}}, \bibinfo {author} {\bibfnamefont {T.}~\bibnamefont {Asaba}},
  \bibinfo {author} {\bibfnamefont {S.}~\bibnamefont {Ikemori}}, \bibinfo
  {author} {\bibfnamefont {Y.}~\bibnamefont {Sekino}}, \bibinfo {author}
  {\bibfnamefont {Y.}~\bibnamefont {Kasahara}}, \bibinfo {author}
  {\bibfnamefont {K.}~\bibnamefont {Totsuka}}, \bibinfo {author} {\bibfnamefont
  {B.}~\bibnamefont {Li}}, \bibinfo {author} {\bibfnamefont {Y.}~\bibnamefont
  {Zhao}}, \bibinfo {author} {\bibfnamefont {Y.}~\bibnamefont {Li}}, \bibinfo
  {author} {\bibfnamefont {Y.}~\bibnamefont {Kohama}},\ and\ \bibinfo {author}
  {\bibfnamefont {Y.}~\bibnamefont {Matsuda}},\ }\href
  {https://arxiv.org/abs/2407.16208} {\bibinfo {title} {Gapless spin
  excitations in a quantum spin liquid state of s=1/2 perfect kagome
  antiferromagnet}} (\bibinfo {year} {2024}{\natexlab{b}}),\ \Eprint
  {https://arxiv.org/abs/2407.16208} {arXiv:2407.16208 [cond-mat.str-el]}
  \BibitemShut {NoStop}%
\bibitem [{\citenamefont {Zheng}\ \emph {et~al.}(2024)\citenamefont {Zheng},
  \citenamefont {Zhang}, \citenamefont {Zhu}, \citenamefont {Chen},
  \citenamefont {Chan}, \citenamefont {Jenkins}, \citenamefont {Kang},
  \citenamefont {Zeng}, \citenamefont {Xu}, \citenamefont {Ratkovski},
  \citenamefont {Blawat}, \citenamefont {Bangura}, \citenamefont {Singleton},
  \citenamefont {Lee}, \citenamefont {Li},\ and\ \citenamefont {Li}}]{YCOHB5}%
  \BibitemOpen
  \bibfield  {author} {\bibinfo {author} {\bibfnamefont {G.}~\bibnamefont
  {Zheng}}, \bibinfo {author} {\bibfnamefont {D.}~\bibnamefont {Zhang}},
  \bibinfo {author} {\bibfnamefont {Y.}~\bibnamefont {Zhu}}, \bibinfo {author}
  {\bibfnamefont {K.-W.}\ \bibnamefont {Chen}}, \bibinfo {author}
  {\bibfnamefont {A.}~\bibnamefont {Chan}}, \bibinfo {author} {\bibfnamefont
  {K.}~\bibnamefont {Jenkins}}, \bibinfo {author} {\bibfnamefont
  {B.}~\bibnamefont {Kang}}, \bibinfo {author} {\bibfnamefont {Z.}~\bibnamefont
  {Zeng}}, \bibinfo {author} {\bibfnamefont {A.}~\bibnamefont {Xu}}, \bibinfo
  {author} {\bibfnamefont {D.}~\bibnamefont {Ratkovski}}, \bibinfo {author}
  {\bibfnamefont {J.}~\bibnamefont {Blawat}}, \bibinfo {author} {\bibfnamefont
  {A.}~\bibnamefont {Bangura}}, \bibinfo {author} {\bibfnamefont
  {J.}~\bibnamefont {Singleton}}, \bibinfo {author} {\bibfnamefont {P.~A.}\
  \bibnamefont {Lee}}, \bibinfo {author} {\bibfnamefont {S.}~\bibnamefont
  {Li}},\ and\ \bibinfo {author} {\bibfnamefont {L.}~\bibnamefont {Li}},\
  }\href {https://arxiv.org/abs/2409.05600} {\bibinfo {title} {Thermodynamic
  evidence of fermionic behavior in the vicinity of one-ninth plateau in a
  kagome antiferromagnet}} (\bibinfo {year} {2024}),\ \Eprint
  {https://arxiv.org/abs/2409.05600} {arXiv:2409.05600 [cond-mat.str-el]}
  \BibitemShut {NoStop}%
\bibitem [{\citenamefont {He}\ \emph {et~al.}(2024)\citenamefont {He},
  \citenamefont {Yu},\ and\ \citenamefont {Li}}]{1-9Z3}%
  \BibitemOpen
  \bibfield  {author} {\bibinfo {author} {\bibfnamefont {L.-W.}\ \bibnamefont
  {He}}, \bibinfo {author} {\bibfnamefont {S.-L.}\ \bibnamefont {Yu}},\ and\
  \bibinfo {author} {\bibfnamefont {J.-X.}\ \bibnamefont {Li}},\ }\bibfield
  {title} {\bibinfo {title} {Variational monte carlo study of the
  $1/9$-magnetization plateau in kagome antiferromagnets},\ }\href
  {https://doi.org/10.1103/PhysRevLett.133.096501} {\bibfield  {journal}
  {\bibinfo  {journal} {Phys. Rev. Lett.}\ }\textbf {\bibinfo {volume} {133}},\
  \bibinfo {pages} {096501} (\bibinfo {year} {2024})}\BibitemShut {NoStop}%
\bibitem [{\citenamefont {Oshikawa}\ \emph {et~al.}(1997)\citenamefont
  {Oshikawa}, \citenamefont {Yamanaka},\ and\ \citenamefont {Affleck}}]{OYA}%
  \BibitemOpen
  \bibfield  {author} {\bibinfo {author} {\bibfnamefont {M.}~\bibnamefont
  {Oshikawa}}, \bibinfo {author} {\bibfnamefont {M.}~\bibnamefont {Yamanaka}},\
  and\ \bibinfo {author} {\bibfnamefont {I.}~\bibnamefont {Affleck}},\
  }\bibfield  {title} {\bibinfo {title} {Magnetization plateaus in spin chains:
  ``haldane gap'' for half-integer spins},\ }\href
  {https://doi.org/10.1103/PhysRevLett.78.1984} {\bibfield  {journal} {\bibinfo
   {journal} {Phys. Rev. Lett.}\ }\textbf {\bibinfo {volume} {78}},\ \bibinfo
  {pages} {1984} (\bibinfo {year} {1997})}\BibitemShut {NoStop}%
\bibitem [{\citenamefont {Furuya}\ and\ \citenamefont
  {Horinouchi}(2019)}]{OYA2D}%
  \BibitemOpen
  \bibfield  {author} {\bibinfo {author} {\bibfnamefont {S.~C.}\ \bibnamefont
  {Furuya}}\ and\ \bibinfo {author} {\bibfnamefont {Y.}~\bibnamefont
  {Horinouchi}},\ }\bibfield  {title} {\bibinfo {title} {Translation
  constraints on quantum phases with twisted boundary conditions},\ }\href
  {https://doi.org/10.1103/PhysRevB.100.174435} {\bibfield  {journal} {\bibinfo
   {journal} {Phys. Rev. B}\ }\textbf {\bibinfo {volume} {100}},\ \bibinfo
  {pages} {174435} (\bibinfo {year} {2019})}\BibitemShut {NoStop}%
\bibitem [{\citenamefont {Furuya}\ \emph {et~al.}(2020)\citenamefont {Furuya},
  \citenamefont {Horinouchi},\ and\ \citenamefont {Momoi}}]{OYAKagome}%
  \BibitemOpen
  \bibfield  {author} {\bibinfo {author} {\bibfnamefont {S.~C.}\ \bibnamefont
  {Furuya}}, \bibinfo {author} {\bibfnamefont {Y.}~\bibnamefont {Horinouchi}},\
  and\ \bibinfo {author} {\bibfnamefont {T.}~\bibnamefont {Momoi}},\ }\href
  {https://arxiv.org/abs/2011.13095} {\bibinfo {title} {Anomalies of kagome
  antiferromagnets on magnetization plateaus}} (\bibinfo {year} {2020}),\
  \Eprint {https://arxiv.org/abs/2011.13095} {arXiv:2011.13095
  [cond-mat.str-el]} \BibitemShut {NoStop}%
\end{thebibliography}%
\setcounter{figure}{0}

\renewcommand{\thefigure}{S\arabic{figure}}
\clearpage
\onecolumngrid
\begin{center}
    {\Large \textbf{Supplemental Materials for ``Valence Bond Crystal Ground State of the 1/9 Magnetization Plateau in the Spin-1/2 Kagome Lattice''}}
\end{center}

The following three figures provide supplementary information that supports the content described in the main text.

\begin{figure}[h]
\includegraphics[width=100mm]{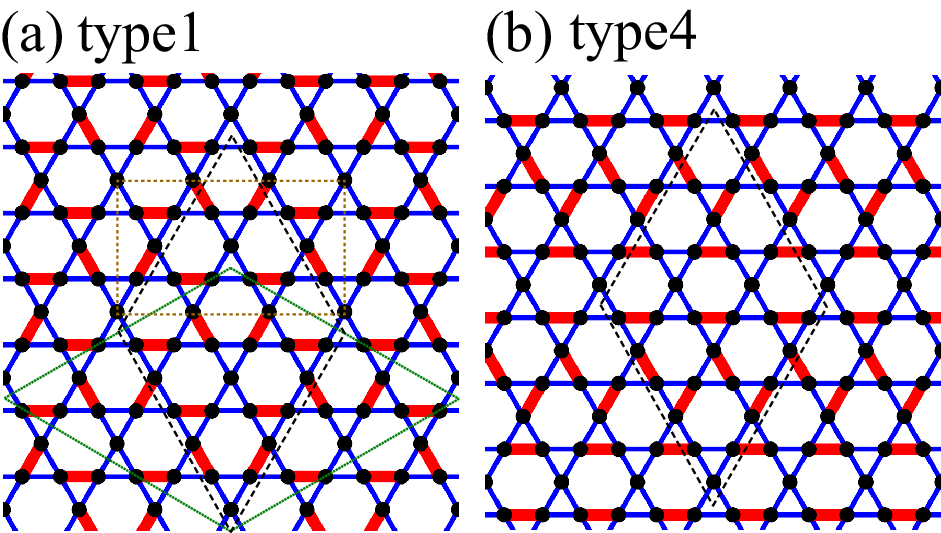}
\caption{Clusters used for exact diagonalization.  (a) the brown rectangle, black rhombus, and green rhombus represent clusters with periodic boundary conditions for $N = 18$, 27, and 36, respectively.  (b) the black rhombus indicates the $N = 27$ cluster with periodic boundary conditions.
}
\end{figure} 

\begin{figure}[h]
\includegraphics[width=86mm]{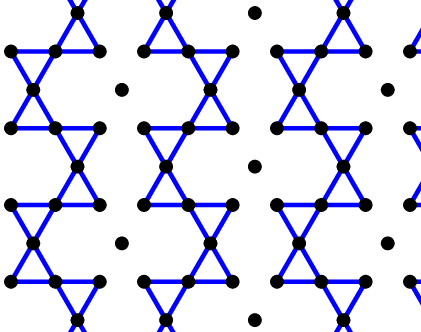}
\caption{
The lattice corresponding to the case where the exchange interactions $J$ belonging to the blue triangles in Fig.~1(a) are set to zero and $J_{\rm d} = J$. In this case, independent $\Delta$-chains are aligned.
}
\end{figure} 

\begin{figure}[h]
\includegraphics[width=86mm]{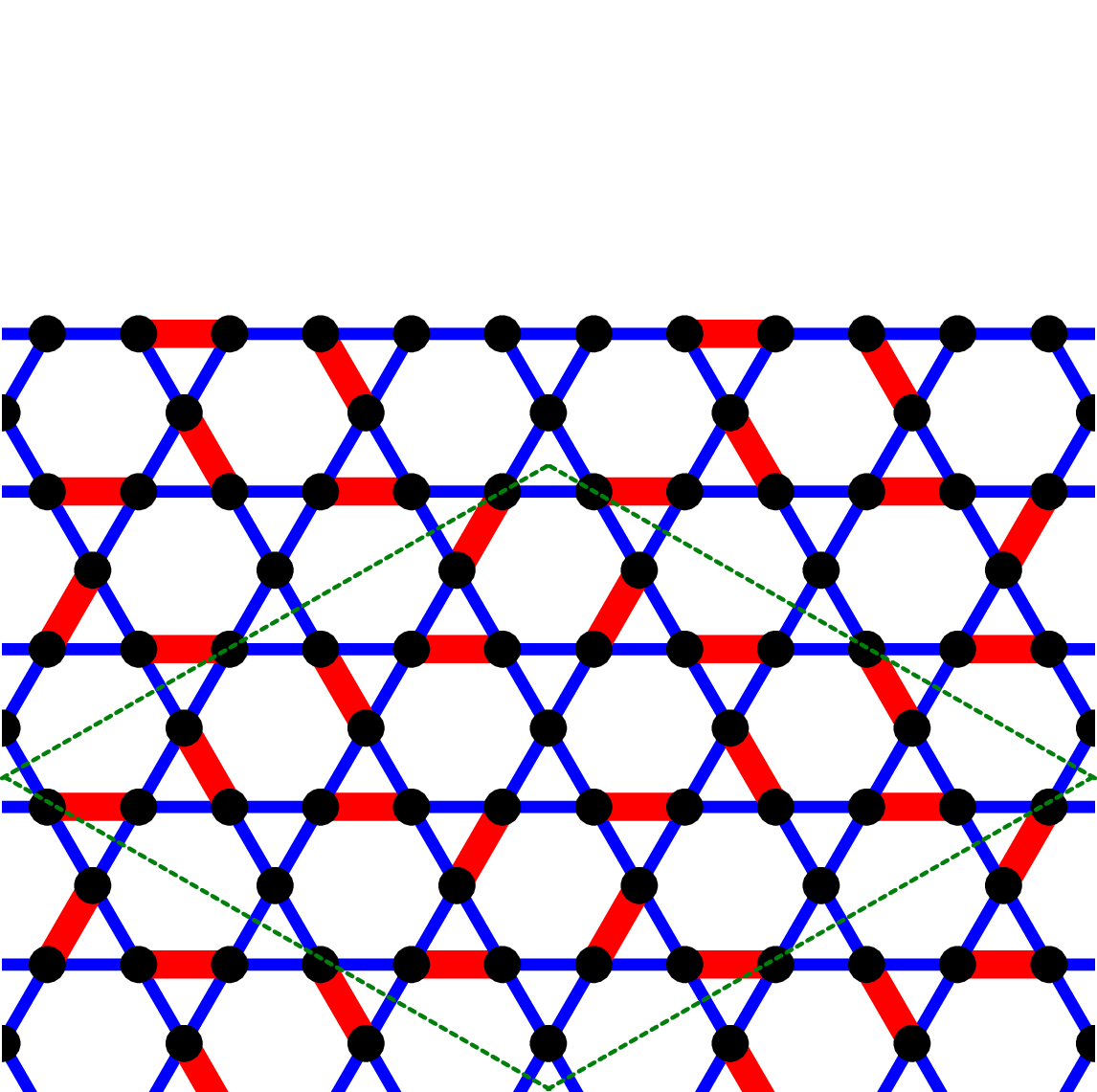}
\caption{The long-period structure of the $N=36$ cluster. The green rhombus represents the cluster with periodic boundary conditions.
}
\end{figure} 

%\bibliography{ref}

%\begin{thebibliography}{99}

%\end{thebibliography}
\end{document}